\documentclass[sigconf]{acmart}
\usepackage[]{collab}
\usepackage{enumitem}
\usepackage{balance}
\usepackage{xcolor}
\usepackage{subfigure}
\usepackage{multirow}
\usepackage[most]{tcolorbox}
\usepackage{tabularx}
\renewcommand\footnotetextcopyrightpermission[1]{} 

\newcommand\param{``<$*$>''\xspace}

\newcommand\cloud{Huawei Cloud\xspace}
\newcommand\nm{ScaleAD\xspace}

\newcommand{\ie}{{\em i.e.},\xspace}
\newcommand{\eg}{{\em e.g.},\xspace}

\definecolor{ballblue}{rgb}{0.13, 0.67, 0.8}

\collabAuthor{jy}{ballblue}{Jinyang Liu}

\collabAuthor{jz}{magenta}{Jiazhen Gu}
\collabAuthor{yt}{blue}{Yintong Huo}
\collabAuthor{zh}{green}{Zhihan Jiang}
\collabAuthor{jj}{purple}{Junjie}

\AtBeginDocument{%
  \providecommand\BibTeX{{%
    \normalfont B\kern-0.5em{\scshape i\kern-0.25em b}\kern-0.8em\TeX}}}

\begin{document}

\title{Log-based Anomaly Detection based on\\ EVT Theory with feedback}



\author{Jinyang Liu}
\author{Junjie Huang}
\affiliation{%
  \institution{The Chinese University of Hong Kong}
  \city{Hong Kong}
  \country{China}}

\author{Yintong Huo}
\author{Zhihan Jiang}
\author{Jiazhen Gu}
\affiliation{%
  \institution{The Chinese University of Hong Kong}
  \city{Hong Kong}
  \country{China}}

\author{Zhuangbin Chen}
\affiliation{%
  \institution{School of Software Engineering, Sun Yat-sen University} \country{China}}

\author{Cong Feng}
\affiliation{%
  \institution{Computing and Networking Innovation Lab,Huawei Cloud Computing Technology Co., Ltd}
  \country{China}}

\author{Minzhi Yan}
\affiliation{%
  \institution{HCC Lab, Huawei Cloud Computing Technology Co., Ltd}
  \country{China}}

\author{Michael R. Lyu}
\affiliation{%
  \institution{The Chinese University of Hong Kong}
  \country{China}}
  


\renewcommand{\shortauthors}{Jinyang Liu, et al.}

\begin{abstract}
System logs play a critical role in maintaining the reliability of software systems. 
Fruitful studies have explored automatic log-based anomaly detection and achieved notable accuracy on benchmark datasets.
However, when applied to large-scale cloud systems, these solutions face limitations due to high resource consumption and lack of adaptability to evolving logs.
In this paper, we present an accurate, lightweight, and adaptive log-based anomaly detection framework, referred to as \texttt{\nm}. 
Our method introduces a Trie-based Detection Agent (TDA) that employs a lightweight, dynamically-growing trie structure for real-time anomaly detection. 
To enhance TDA's accuracy in response to evolving log data, we enable it to receive feedback from experts.
Interestingly, our findings suggest that contemporary large language models, such as ChatGPT, can provide feedback with a level of consistency comparable to human experts, which can potentially reduce manual verification efforts.
We extensively evaluate \nm on two public datasets and an industrial dataset. The results show that \nm outperforms all baseline methods in terms of effectiveness, runs 2$\times$ to 10$\times$ faster and only consumes 5\% to 41\% of the memory resource.

\end{abstract}

\maketitle

\section{Introduction}
\noindent Cloud computing has surged into popularity in recent years.
Large-scale cloud vendors, such as AWS, Azure, and GCP, provide 7$\times$24 services to customers over the world. 
Ensuring the reliability of cloud systems is a critical task~\cite{chen2020incidental,gunawi2016does,chen2020towards}, because a small period of downtime could result in significant financial loss for both cloud vendors and their customers~\cite{chen2019empirical}.
A preliminary step to safeguard reliability is timely and accurate detection of suspicious system behaviors, \ie anomaly detection. 
Similar to traditional software systems, logs in cloud systems provide valuable insights into the system's functioning and potential issues. 
Log-based anomaly detection, \ie timely identifying anomalous log messages for prompt resolution of issues, has been widely recognized as an essential task of cloud system management~\cite{he2016experience,chen2021experience,zhu2019tools,he2018identifying, he2020loghub}.


Existing approaches typically adopt machine learning or deep learning-based techniques to identify anomalous logs. 
Traditional machine learning-based approaches primarily consider statistical information (\eg log occurrence counts) and apply models such as isolation forest (IF)~\cite{liu2008isolation}, support vector machine (SVM)~\cite{liang2007failure}, decision tree (DT)~\cite{chen2004failure}, logistic regression (LR)~\cite{bodik2010fingerprinting}, to identify anomalies. 
Besides, recent studies have explored the use of deep learning methods to process logs. 
Such approaches typically extract semantic information from log messages through word embedding~\cite{zhang2019robust, yang2021plelog} or Bidirectional Encoder Representations from Transformers (BERT)~\cite{le2021log}, and perform anomaly detection accordingly.

Although previous studies have demonstrated impressive performance on benchmark datasets, they are not practical for production cloud systems due to the following two reasons.
First, previous solutions tend to pursue a high detection accuracy while overlooking the optimization of computation and space complexity.
Cloud systems often use instances (\ie virtual servers) to host customers' applications.
Conducting log anomaly detection for each instance enables close monitoring of its health status. 
However, as customers' applications already take up most resources of the instances, there are limited resources left to run an anomaly detector, which renders existing solutions impractical.
For instance, deep learning-based methods~\cite{zhang2019robust, le2021log} require heterogeneous accelerators (\eg GPU and FPGA) \cite{mohan2022looking, weng2022mlaas} to allow real-time inference, which may not be available to every instance. 
A straightforward approach is to transmit the log data to centralized compute nodes with abundant computational resources, which subsequently return the detection results.
However, instances in a cloud can produce extensive amounts of log data (\eg Azure reported that 5 billion log messages are generated per day \cite{wang2022spine}), that are distributed across different clusters and datacenters. 
Transmitting such large-scale distributed logs to compute nodes could cause additional network and I/O overhead, which is prohibitively expensive and time-consuming.

The second reason is that previous approaches struggle to be adaptive to deal with diverse and evolving log data. 
In cloud systems, frequent launches of new versions of software result in changes to logging statements over time. 
For instance, Google has reported that there are thousands of newly-added logging statements due to software updates every month~\cite{xu2010system}. 
Existing methods typically train models to learn anomaly patterns from historical log data, which can hardly adjust to new logs, causing performance degradation.
To address this problem, a recent study~\cite{zhang2019robust} exploits the semantic similarity between historical logs and new logs, enabling the algorithm to transfer knowledge about anomalies from historical data to new data for anomaly identification.
However, logs in real systems are complicated, and whether the assumption (\ie new logs share similar semantics with the historical ones) holds in real cloud systems has not been well investigated.


In this paper, we conduct a study on the logs in \cloud (a large-scale cloud system) to better understand the characteristics of logs in production cloud systems. 
We observe that an instance could generate several gigabytes of logs daily, making log anomaly detection critical for it. 
However, only very limited resources (\eg one CPU core and 200MB memory) are left for a plug-in anomaly detection process.
Besides, logs are evolving and have low semantic similarities. 
Our study finds that during a one-month-long development cycle of 20 microservices, approximately 14.5\% of brand-new logs are introduced on average.
When comparing the semantic similarities between two versions of the same microservice, we find that 85\% of log message pairs share little semantic similarity. 

Based on the above results, we can summarize that a practical log-based anomaly detection method for cloud systems should be {\em accurate}, {\em lightweight}, and {\em adaptive}. 
It is non-trivial to achieve all the above three requirements simultaneously. 
Lightweight anomaly detection methods (\eg logistic regression~\cite{bodik2010fingerprinting}) cannot effectively handle newly occurring anomalous logs (\ie not adaptive). 
On the other hand, adaptive methods (\eg RobustLog~\cite{zhang2019robust}) need to extract semantic information from logs, which is compute-intensive (\ie not lightweight). 
It is challenging, if not infeasible, for pure data-driven methods to be both lightweight and adaptive.
In practical scenarios, the involvement of engineers is crucial for system operation~\cite{chen2020towards,wang2021fast,ghosh2022fight}.
In \cloud, on-call engineers must manually confirm the reported anomalies before initiating the mitigation process.
Based on this observation, we propose to incorporate the valuable feedback (\ie the ground-truth labels of anomalies) from experts, which is necessary for a practical (\ie both lightweight and adaptive) log-based anomaly detection method in cloud systems.

Based on such intuition, we propose \texttt{\nm}, a scalable and adaptive log-based anomaly detection framework with experts in the loop for cloud systems.
To meet the lightweight and adaptiveness requirement, \nm employs a trie-based detection agent (TDA) to efficiently perform anomaly detection.
TDA utilizes a lightweight trie structure (also known as a prefix tree) that stores received log messages in a compact form by allowing log messages to share the same internal nodes.
In addition, the trie structure can be dynamically expanded during runtime to incorporate new logs via incrementally adding new nodes to the trie.
Using the trie structure, we can efficiently calculate the occurrence distribution of different logs, based on which we apply extreme value theory (EVT)~\cite{siffer2017anomaly}, an efficient statistics-based method, for anomaly detection. 

Our TDA is lightweight and can be locally deployed on the instance of interest without consuming excessive resources. It raises an alarm to experts (such as on-call engineers) for verification when anomalous logs are detected.
TDA then incorporates the verification results as feedback to continuously improve its accuracy. 
To reduce the workload of manual verification for on-call engineers, we allow various forms of experts. 
One option is to use a knowledge base storing rules obtained from historical operations, such as keyword matching.
Besides, inspired by the recent success of large language model (LLM)~\cite{min2021recent} (\eg ChatGPT~\cite{chatgpt}), we propose that leveraging LLMs as an expert is a promising approach. 
By leveraging the LLM's ability to comprehend the semantic meaning of logs, we can determine whether they are anomalous or not. Our case study, which is based on real-world industrial logs in \cloud, demonstrates that ChatGPT is capable of providing feedback that is consistent with that of on-call engineers to some extent.

To evaluate the performance of \nm, we conduct extensive experiments on two widely-used public datasets and an industrial dataset collected from \cloud.
The experimental results demonstrate that compared with state-of-the-art solutions, \nm achieves the best performance (0.908 to 0.990 F1 scores) in the setting where log data is fixed.
\nm retains a high and stable accuracy even if log messages evolve. Moreover, compared with existing methods, \nm is 2$\times$ to 10$\times$ faster and consumes 5\% to 41\% fewer memories.

To sum up, this paper makes the following contributions:
\begin{itemize}[leftmargin=*, topsep=0pt]
    \item We conduct a study to understand the practical requirements for log-based anomaly detection methods in cloud systems (Section~\ref{sec: empirical_study}). The study revealed that a practical method should be accurate, lightweight, and adaptive, which aims to bridge the gap between research on log-based anomaly detection and its real-world application.
    \item We propose \nm, a novel log-based anomaly detection framework. It comprises a lightweight trie-based detection agent and an expert module in the loop to achieve all three practical requirements, \ie accurate, lightweight, and adaptive (Section~\ref{sec: method}).
    \item We conduct extensive evaluations of \nm on both public and industrial datasets. The results demonstrate that \nm outperforms state-of-the-art methods in terms of effectiveness, adaptability, efficiency, and space consumption (Section~\ref{sec: evaluation}).
\end{itemize}
\section{Background and Motivation}
\label{sec: background}

\noindent In this section, we first discuss the background of log-based anomaly detection.
Then, we present a study to comprehend the characteristics of logs within \cloud. Based on this understanding, we outline the industrial requirements that inspire our method design.


\subsection{Log-based Anomaly Detection}
\noindent The goal of log-based anomaly detection is to identify the log messages that may indicate a system problem in the runtime.
In the literature, most studies adopt a two-step paradigm \ie \textit{log parsing} and \textit{anomaly detection}~\cite{he2016experience,chen2021experience,le2022log}. 
In the log parsing step, these methods obtain log templates via identifying the constant and variable parts.
For instance, consider the raw log message ``Finished task 0.0 in stage 6.0 (TID 247).'' Its log template would be ``Finished task <$*$> in stage <$*$> (TID <$*$>),'' where the parameters (0.0, 6.0, and 247) are replaced with ``<$*$>.''
The sequence of log templates is then processed through various downstream methods to conduct anomaly detection. For example, the library Loglizer~\cite{he2016experience} applies various machine learning-based methods (\eg logistic regression and decision tree) to detect anomalies based on the count distribution characteristics of templates.

Although log-based anomaly detection methods have long been recognized as an important problem for software maintenance, their practical application in cloud scenarios is still understudied.
Unlike traditional software, cloud systems operate on a larger scale, generating massive volumes of logs from millions of physical and virtual instances.
This presents new requirements for log analysis, which we aim to investigate in the following section by conducting a study on log data characteristics in cloud systems such as \cloud.

\subsection{Characteristics of Log Data in \cloud}\label{sec: empirical_study}
\noindent To study the characteristics of log data in \cloud, we collect one-month log data from 20 internal microservices in our production environment.
These microservices are from the IaaS layer of \cloud, spanning functionalities such as API Gateway, user management, auto-scaling, and load balance.
In the following study, we aim to understand the practical requirements for log-based anomaly detection methods in cloud systems.

\subsubsection{Massive and Distributed Log Data.}
\label{sec: massive_diverse_log}

The huge volume of logs generated by modern cloud systems has been widely recognized in recent studies~\cite{wang2022spine, liu2019logzip}. 
In \cloud, a single service can produce tens of gigabytes of log data per day due to the frequent invocation by a large number of users. Moreover, this large volume of log data is distributed on different instances, resulting in great challenges for log analysis.
A direct solution for log analysis is to transmit the log data to a central service for analysis, which, however, is not practical given the associated transportation cost in terms of time and bandwidth. 
Therefore, analyzing log data locally within the instance is desirable. 
However, in multi-tenant cloud systems, vendors pack resources such as CPU cores and memory as instances (\eg virtual machines) that are provided to external or internal users to run various applications.
The running application on an instance usually can consume most of the resources, leaving very limited resources for plug-in processes such as log-based anomaly detection.
For example, in \cloud, we find that most control plane instances have only $\sim$200MB memory and a single CPU core that can be allocated to an anomaly detector without adversely affecting other applications on the instance. 

Recent state-of-the-art solutions using deep learning-based techniques, (\eg LSTM~\cite{du2017deeplog, meng2019loganomaly, zhang2019robust} and transformers\cite{le2021log}), achieves promising anomaly detection accuracy.
However, these methods require significant memory resources to store a large number of parameters, such as word embeddings, and their inference efficiency is inadequate without GPU acceleration, as demonstrated in Section~\ref{sec: exp_time_space}. 
In conclusion, a lightweight method that can quickly process log analysis without consuming too much memory is essential to address the challenges posed by massive and distributed log data.

\begin{figure}[t]
  \centering
  \mbox{
     \subfigure[Templates Jaccard similarity  of 20 Microservices\label{fig: jccard_similarity}]{\includegraphics[width=0.465\columnwidth]{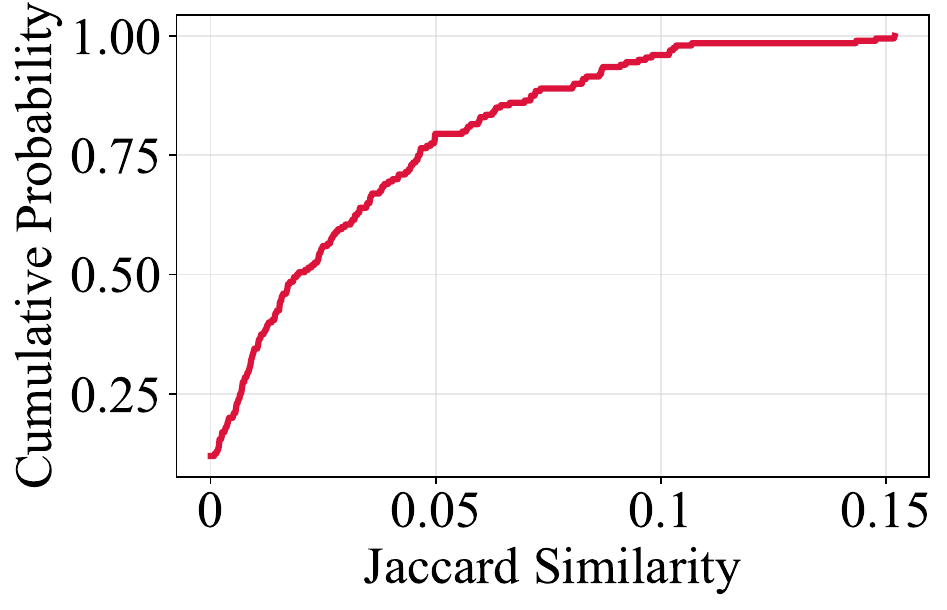}}\quad
    
    \subfigure[Semantic similarity comparison of log templates between two versions of the same microservices 
    \label{fig: semantic_similarity}]
    {\includegraphics[width=0.465\columnwidth]{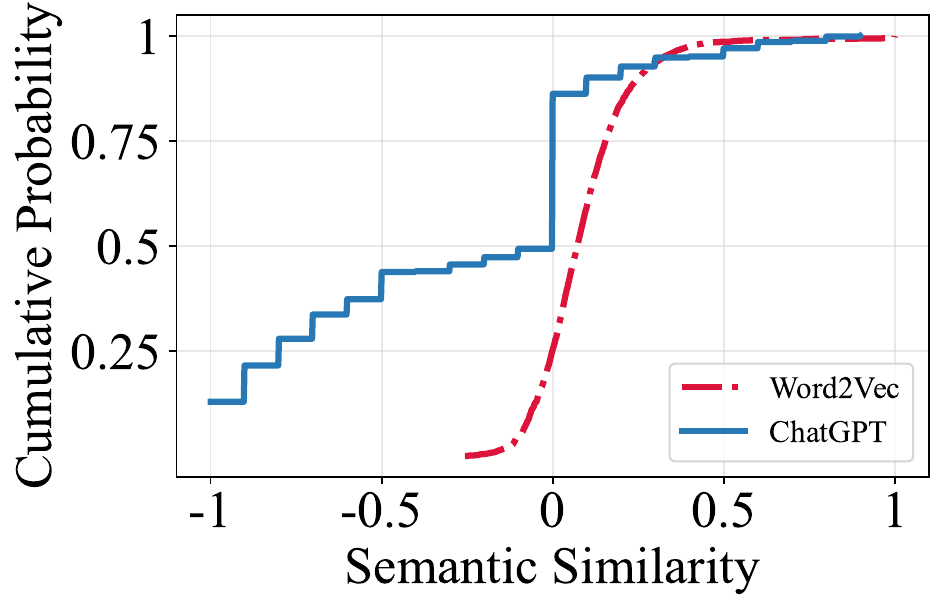}}
    }
    \vspace{-12pt}
  \caption{Statistics of log messages in \cloud}
  \vspace{-12pt}
\end{figure}

\begin{figure*}[t]
    \centering
    \includegraphics[width=2\columnwidth]{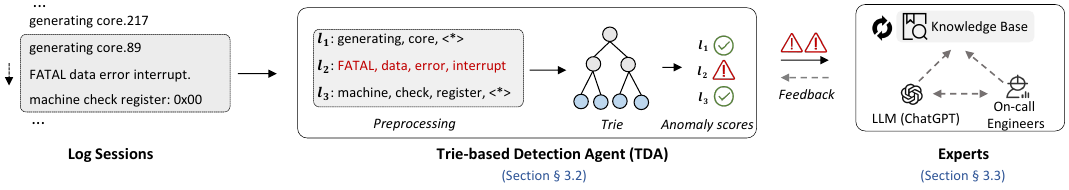}
    \vspace{-15pt}
    \caption{The overall framework of \nm}
    \label{fig: overall_framework}
\vspace{-6pt}
\end{figure*}

\subsubsection{Evolving and Diverse Log Data.}
\label{sec: evolving_log}
Besides the huge log volume, the current software development process typically adopts the continuous integration/delivery (CI/DI) paradigm~\cite{vassallo2019automated}, making changes to software and deployments increasingly frequent.
As a result, the log data in cloud systems change over time, \eg modifications to logging statements and new-added log messages, causing great challenges to downstream log analysis tasks.

To better understand the situation of log evolution in \cloud, we study the number of log template changes from 01/02/2023 to 01/03/2023 of 20 microservices. These templates are obtained by running a log parser Drain~\cite{he2017drain} as previous studies~\cite{zhang2019robust, wang2022spine}.
Our analysis reveals that a one-month evolution in \cloud can lead to a 14.5\% increase in the number of new log templates. Such changes are typically due to actions like fixing bugs and developing new features. Our findings indicate that log template evolution is a natural and continuous process in \cloud.

Besides the change of template numbers, we further study how its content changes.
First, we enumerate each pair of microservices among the selected 20 microservices and measure how their log templates overlap.
Specifically, we denote the template sets of two microservice as $set_i$ and $set_j$, each of which contains unique templates of corresponding 
microservices. Then we compute the Jaccard similarity, \ie $J(set_i,set_j) = \frac{|set_i \cap set_j|}{|set_i \cup set_j|}$, where $|\cdot|$ denotes the number of unique template in a set.
We plot the CDF (cumulative distribution function) figure of the similarity distribution in Figure~\ref{fig: jccard_similarity}. We find that around 97\% of pairs of microservices have less than 0.1 Jaccard similarities in their template sets. This is intuitive since different microservices tend to log different information. 

Second, we measure the extent to which newly-added logs share semantics with historical logs when a microservice undergoes an upgrade. To accomplish this, we selected the most frequently updated microservice and compared the semantic similarity between its log messages before and after the update.
We computed the semantic similarity for each pair of historical and new log messages by using Word2Vec~\cite{mikolov2013efficient} to obtain a semantic vector for each log and computed a cosine similarity to represent the semantic similarity between each pair.
The cosine similarity ranges from -1 (most dissimilar) to 1 (most similar).
Furthermore, to obtain more reliable results, we used ChatGPT, a language model known for its ability to understand text semantics, to score the semantic similarity of each log message pair on a scale of -1 to 1.
We plotted the CDF of similarities of these log pairs in Figure~\ref{fig: semantic_similarity}. Our findings reveal that Word2Vec and ChatGPT produce similar results, indicating that most log pairs ($\sim 95\%$) have similarities less than 0.3. Notably, ChatGPT regards 85\% of log pairs as having little semantic similarity ($\leq 0$). These results suggest that software updates can introduce new log messages that differ significantly from historical logs.

\citet{zhang2019robust} attempts to tackle the log evolving issue by ensuring robust representation of log data. The assumption is that different log messages could share similar semantics (such as synonyms). However, our study has demonstrated that this is hardly applicable, given the significant diversity of logs, even within the same microservice.
In conclusion, there is a practical need for a method that can adaptively and accurately detect anomalies from the evolving and diverse log data in real-world scenarios.

\subsection{Demanding Industrial Requirements}\label{sec: industrial_requirements}
\noindent We summarize the following industrial requirements based on our study in \cloud as introduced in Section~\ref{sec: empirical_study}.

\begin{itemize}[leftmargin=*, topsep=0pt]
    \item \textbf{Accurate}. High accuracy is crucial for an anomaly detector, as it indicates that the detector can identify potential issues in logs without generating too many false positives, which may distract on-call engineers. This is the primary priority of a practical log-based anomaly detector.
    \item \textbf{Lightweight}. To process large amounts of log data on instances with limited computational capability and memory without affecting running applications, a lightweight anomaly detector should (1) use minimal memory resources and (2) efficiently handle massive streaming log messages, preferably without the acceleration of special devices (e.g., GPU).
    \item \textbf{Adaptive}. A practical method for log-based anomaly detection should be able to learn adaptively from diverse microservices log data, incorporate human feedback to address false positives and maintain accuracy, and operate continuously in an online production environment without interruption, such as retraining and redeployment.
\end{itemize}

\section{Method}\label{sec: method}
\noindent In this section, we present the proposed framework, \nm, designed to address the stringent industrial requirements. We begin with an overview of \nm in Section~\ref{sec: method_overview}, followed by a detailed explanation of its components in the subsequent sections.

\subsection{Overview}\label{sec: method_overview}

\nm comprises two core elements:
a trie-based detection agent (TDA) and an expert module, as illustrated in Figure~\ref{fig: overall_framework}.
We first collect streaming log messages generated by an instance and arrange them into sessions (or windows). Next, TDA receives the windows of log messages as input and produces an anomaly score for each log line within the window.
During this process, TDA identifies any suspicious logs and forwards them to experts for further verification. The experts then return their feedback to TDA, which integrates this feedback to continuously refine its accuracy. 

Within this framework, TDA utilizes a lightweight trie structure to parse log messages into templates and detect anomalies based on the count distribution of those templates. The trie structure stores a compact record of encountered log templates and their corresponding counts to ensure efficient processing of high-volume log data.
Moreover, the trie structure is designed to be adaptive and can be dynamically expanded during runtime to accommodate new templates. 
Due to its efficiency in design, TDA can be deployed locally on an instance, while we allow a remote expert to provide feedback to TDA. 
The expert could take many forms, such as an on-call engineer, a knowledge base containing various rules, or a computational-intensive large language model (LLM). 
By leveraging the expertise of a remote specialist, TDA can further improve its accuracy and reduce the likelihood of false positives.



\subsection{Trie-based Detection Agent}\label{sec: Trie-based Detection Agent}
Trie-based Detection Agent (TDA) processes raw log messages as input and generates an anomaly score for each log message as output. 
The TDA comprises five steps: \textit{preprocessing}, \textit{internal node traverse}, \textit{leaf node update}, \textit{trie update}, and \textit{anomaly detection}.
At the heart of the TDA is a trie, consisting of \textit{internal nodes} and \textit{leaf nodes}. 
Log messages traverse the trie based on heuristics defined at internal nodes, which take into account various log message characteristics. The objective is to assign similar log messages to the same leaf node in a coarse-grained manner.
Upon reaching a leaf node, a log message is assigned to a \textit{log cluster}. Each log cluster comprises log messages sharing the same template. Finally, anomaly detection is carried out efficiently using the statistical information gathered from the log clusters.
In the following, we provide a detailed explanation of each step within the TDA.

\subsubsection{Preprocess}
For an input log message $l_i$, we follow a common practice~\cite{wang2022spine, zhu2019tools} to use pre-defined regular expressions to extract parameter fields such as IP address and URL. Note that one can also apply other regular expressions according to logs produced by a specific system. 
After that, we tokenize the log message with non-alphanumeric splitters (\ie any characters that are not letters or numbers), generating log tokens $\hat{l}_i=[w_1, w_2, ...]$. 
In the following, we consider each log message comes in a streaming manner.

\subsubsection{Internal node traverse}
We defined the following three general heuristics within internal nodes of the Trie, which can be generally applied to log messages generated by various systems. Figure~\ref{fig: trie} presents a running example of Trie.

\textit{(1) Traverse by domain knowledge.}
Initially, log messages are separated based on distinguishing features, such as levels (e.g., INFO and DEBUG) and components, since log messages with different levels or generated by different components are more likely to belong to separate templates.

\textit{(2) Traverse by most frequent tokens.}
As frequently occurring tokens in log messages are more likely to be constant parts of a template~\cite{liu2019logzip}, we extract the $K$ most frequent tokens from each log message's tokens $l_i$ as a \textit{token key}. Log messages with different token keys are then separated. Generally, $K$ is set to 3, but may be increased for systems producing lengthy log messages. \nm maintains a vocabulary that counts token occurrences while processing log messages. English stopwords are discarded when extracting token keys to avoid grouping dissimilar log messages.

\textit{(3) Traverse by prefix tokens.} Inspired by the observation of Drain~\cite{he2017drain} that tokens at the beginning of a log message are more likely to be constants,
we divide log messages based on their first $d$ prefix tokens one by one, \eg ``open'' and ``file'' in the example of Figure~\ref{fig: trie}, 
Generally, setting $d$ to 3 yields satisfactory results. 
However, prefix tokens may sometimes be parameters, resulting in many child nodes in the trie. 
To address this issue, we use a hyper-parameter $c_{max}$ to limit the maximum number of child nodes.
For example, when setting $c_{max}=3$, the 4th child of a leading token will be replaced by a parameter \param that can match any tokens.

\subsubsection{Leaf node update}~\label{sec: leaf_node_update}
Upon traversing all internal nodes in the Trie, a log message $l_i$ will eventually reach a leaf node.  
Leaf node update is then performed to extract a log template for each log.
In particular, each leaf node contains multiple \textit{log clusters}, represented as tuples of the form $C_i = (\mathbf{L_i}, t_i)$, where $\mathbf{L_i}$ records the IDs of all log messages in the cluster, and $t_i$ denotes the template shared by these log messages.

\begin{figure}[t]
    \centering
    \includegraphics[width=0.95\columnwidth]{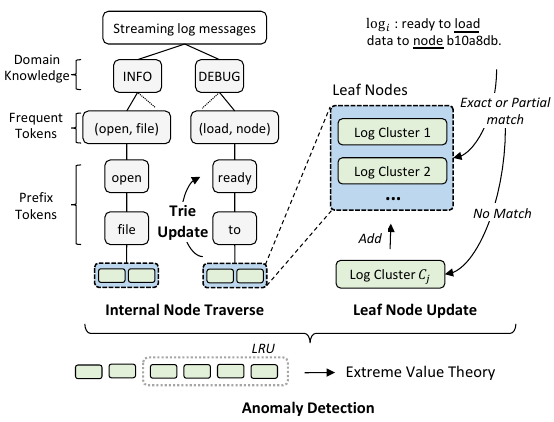}
    \vspace{-12pt}
    \caption{The workflow of trie-based detection agent (TDA)}
    \label{fig: trie}
\vspace{-18pt}
\end{figure}

As shown in the right side of Figure~\ref{fig: trie}, we obtain the template of the arriving log by matching it with a log cluster in the leaf node.
We first attempt to match it with existing log clusters (\ie exact or partial match). If this matching process fails (\ie no match), we create a new log cluster containing only the arriving log message. The associated template of the log cluster is then updated based on the arriving log message.
We elaborate on exact match, partial match and not match as follows.

\textit{(1) Exact match.} 
Firstly, we regard the template of each log cluster as a regular expression (e.g., replace <*>'' as .*?'') to match the given log message. This approach allows the template to match longer log messages, even if they contain varying-length parameters, as the regular expression allows matching tokens of any length. For example, the regular expression ``dumping .*?'' can successfully match ``dumping item1, item2'' and ``dumping item0''.

\textit{(2) Partial match.}
Next, if the exact match fails, we conduct a partial match. Specifically, we tokenize the template and calculate the Jaccard similarity between the template and the candidate log message. We denote the tokenized template as $\hat{t}_i=[w_1, w_2, ...]$. The Jaccard similarity $J(\hat{l}_i,\hat{t}_i)$ can be calculated as $J(\hat{l}_i,\hat{t}_i) = (|\hat{l}_i \cap \hat{t}_i|) / (|\hat{l}_i \cup \hat{t}_i|)$, where $|\cdot|$ denotes the number of unique tokens in a list. We calculate the distance between $\hat{l}_i$ and each template of existing log clusters.
If the \textit{maximum} distance $\hat{d}_J(\hat{l}_i,\hat{t}_i)$ is larger than a threshold $\theta_{match}$ (typically set to 0.5), $l_i$ can match $t_i$, otherwise, it cannot match any existing log clusters (\ie no match).

\textit{(3) No match.}
If no log clusters in the leaf node can match the given log message, we create a new log cluster (e.g., $C_j=(L_j, t_j)$) in the leaf node. The template for this new log cluster is the log message itself, \ie $t_j=l_i$, and this log cluster contains only one log message, \ie $L_j = [i]$.

\textit{Template update.}
If the given log message $l_i$ matches a log cluster $C_i$, we add the identifier $i$ to the record $L_i$ of $C_i$, and update the template $t_i$ for the matched log cluster using $l_i$. Specifically, we first identify the common set of tokens shared by both $t_i$ and $l_i$. Next, we choose the list that has more tokens between $\hat{t}_i$ and $\hat{l}_i$. Finally, we replace any token in the longger list that is not in the common token set with the placeholder \param. This process generates the new template for the log cluster $C_i$.

\subsubsection{Trie update}~\label{sec: trie_update}
Diverse log messages received in a continuous stream can result in inaccurate template extraction since false templates can be produced before a sufficient number of log messages accumulate.
This, in turn, affects the accuracy of both exact and partial matching steps, 
producing more false log clusters.
For instance, two log clusters with similar templates may coexist: $t_1$ as ``deleted items: <*>'' and $t_2$ as ``deleted items: obj1, obj2, obj3, obj4''.
The reason is that $t_2$ has many parameters (obj1-obj4), leading to higher Jaccard distance and cannot be merged with $t_1$ when doing leaf node update (\ie no match). 

To tackle this problem, we propose a bottom-up Trie update approach that merges log clusters sharing the same template into a single cluster and reconstructs the Trie based on the new template of the merged log cluster. The approach starts by identifying a set of log cluster candidates denoted as $S_c=[C_1, C_2, ... C_M]$
that share the same "domain knowledge" and "frequent token" internal nodes.
Candidates are then sorted based on the number of \param in their templates, and those with more parameters are prioritized. The approach iterates through each candidate template and checks if the preceding template \textit{contains} the succeeding one. If so, the succeeding log cluster is deleted, and its log sequences are merged with the preceding one.
To determine if template A \textit{contains} template B, we generate a regular expression from A (by replacing \param with ".*?") and verify its match with template B. Finally, the Trie is constructed using the remaining log templates, yielding a more concise and accurate trie with a reduced template count.

Updating trie excessively for each log message will obfuscate TDA's efficiency, as it involves matching regular expressions between existing templates and reconstructing the Trie. To optimize efficiency while ensuring Trie accuracy, we trigger Trie updates periodically, such as every million lines of log messages.

\subsubsection{Anomaly detection}\label{sec: anomaly detection}
Log messages are parsed to templates after traversing the Trie structure.
In previous approaches~\cite{he2016experience, meng2019loganomaly}, feature extraction and anomaly detection are then performed separately, leading to additional computation and potential maintenance costs.
To address this issue, we propose a lightweight anomaly detection module that integrates seamlessly with the Trie.

Our approach aims to identify anomalies by detecting templates that appear significantly less frequently than others. To achieve this, we leverage the Generalized Extreme Value (GEV) distribution within the Extreme Value Theory (EVT) framework~\cite{siffer2017anomaly}, which can efficiently identify extreme values within a large set of values.
Specifically, we enumerate each log cluster, count the occurrences of each template, and generate the count list $L_{count}=[x_1, x_2, ..., x_R]$, where $x_i$ represents the count of the $i_{th}$ template, and $R$ is the number of templates considered.
Next, we apply the GEV to detect anomalies in $L_{count}$ by fitting the GEV distribution as follows:
\begin{equation}\label{equ: gev_distribution}
f(x|\mu, \sigma, \xi) = \frac{1}{\sigma}\left[1 + \xi\left(\frac{x-\mu}{\sigma}\right)\right]^{-\frac{1}{\xi}},
\end{equation}
where $f(\cdot)$ represents the cumulative distribution function (CDF) of GEV, $x$ is variable, and $\mu$, $\sigma$, and $\xi$ are the location, scale, and shape parameters of the distribution, respectively, 
We choose EVT because (1) it enables the efficient estimation of GEV parameters by considering only extreme values (such as minimal values). (2) the GEV distribution requires only three parameters ($\mu$, $\sigma$, and $\xi$), making it memory-friendly.
We proceed to compute an anomaly score $tp$ for each log template, given by the following equation:
\begin{equation}\label{equ: anomaly_score}
tp = \frac{f(x)^{\tau}}{\sum_{j=1}^{R} f(x_j)^{\tau}}.
\end{equation}
In order to obtain a softer distribution of anomaly scores, inspired by Hinton et al~\cite{hinton2015distilling}, we introduce a temperature parameter $\tau$ into the formula. A lower $\tau$ value produces softer scores that are more evenly distributed across templates, while a higher $\tau$ value generates more extreme scores that concentrate heavily on a smaller subset of templates. We find that setting $\tau=10$ leads to a satisfactory distribution of anomaly scores in practice.

When dealing with a large number of log templates, the entire set of log clusters can be computationally expensive when fitting the GEV distribution. To address this issue, we propose using the Least-Recently-Used (LRU) strategy to limit the number of log templates considered for fitting the distribution.
The LRU strategy maintains a cache of the most recently seen log templates and discards the least recently used templates when the size of the cache reaches a user-defined size $R$ (as used in Equation~\ref{equ: anomaly_score}).
The LRU strategy is based on the observation that only a small proportion of log messages are typically generated within a short time interval, such as ten minutes. Therefore, we can focus on the most recently seen log templates, which are likely to be relevant for anomaly detection during that time interval.

\subsection{Leveraging Expert Feedback}\label{sec: expert}
\noindent Log evolution can inevitably lead to false positives when applying the TDA. 
For instance, when a new log template appears in a log window, its occurrence count is one, generating a high anomaly score. It is challenging to determine whether this truly indicates an anomalous log message.
While semantic-based methods such as RobustLog~\cite{zhang2019robust} may address such cases by comparing the semantics with seen logs in training data, the semantics among different log messages usually exhibit significant differences (as shown in Section~\ref{sec: massive_diverse_log}), rendering these solutions ineffective.

To address this issue, we propose to incorporate the knowledge of experts to facilitate the anomaly detection process, \ie ground-truth labels for log messages.
It is a common practice in cloud systems that on-call engineers must manually verify every reported suspicious anomaly before mitigating a problem~\cite{wang2021fast,chen2020towards}.
This inspires us to seamlessly integrate such valuable feedback to \nm.
Specifically, when TDA identifies a suspicious anomaly, it issues a query to experts for confirmation.
We use a querying threshold $\theta_{query}$ to control the number of queries issued. Only log templates with an anomaly score $tp$ greater than $\theta_{query}$ are sent to an expert.
Upon receiving a log template, an expert provides feedback in the form of a tuple ($decision$, $ep_i$), where $decision=1$ indicates that the log template represents an anomaly, while $decision=0$ indicates a normal log message. The expert also assigns a confidence score $ep$, which ranges from 0 to 1.
Then we compute an integrated anomaly score $p$, which is a weighted average of TDA's output and the expert's feedback,  where we use the expert's confidence as the weight. 
Formally:
\begin{equation}
p = \begin{cases}
ep + (1 - ep) \times tp, & \text{if } decision = 1\\
1 - \left(ep + (1 - ep)\times(1 - tp) \right), & \text{if } decision = 0\
\end{cases}
\end{equation}
To enable TDA to retain the expert's feedback, we store the decision, i.e. ($decision$, $ep_i$), within the corresponding log clusters in TDA after a log template is confirmed. This enables TDA to make the same decision guided by the expert in the future.

The experts can take multiple forms. \textit{On-call engineers} with substantial domain knowledge can usually provide reliable feedback. However, too many queries from TDA could be distracting and introduce intensive manual efforts. Besides minimizing the query numbers by controlling $\theta_{query}$, we find \textit{Large language models (LLM)} such as ChatGPT could work as a promising expert (will show in Section~\ref{sec: case_study}). 
LLMs possess significant capabilities in understanding the semantics of natural languages, which enables them to identify potential errors described in logs.
As depicted in Figure~\ref{fig: overall_framework}, the feedback obtained from experts is continuously cached to a centralized knowledge base that is shared among the different TDAs distributed across different instances. This ensures that redundant queries are avoided, making \nm execute more efficiently.
In addition, EVT only detects a small portion of anomalous log templates. Consequently, the number of queries to experts is considerably lower than the total amount of log data (will show in Section~\ref{sec: exp_expert_feedback}). As a result, sending queries does not occupy an excessive amount of network bandwidth.

\section{Evaluation}\label{sec: evaluation}
\noindent We evaluate \nm by answering the research questions (RQs).
\begin{itemize}[leftmargin=*, topsep=0pt]
    \item RQ1: What is the effectiveness of \nm under the \textit{offline} setting?
    \item RQ2: What is the effectiveness of \nm under the \textit{online} setting?
    \item RQ3: How experts' feedback affects the performance of \nm?
    \item RQ4: What is the time and space efficiency of \nm?
\end{itemize}



\subsection{Evaluation Setting}
\subsubsection{Dataset}
We evaluate \nm on two widely-used public datasets and an industrial dataset gathered from the production environment of \cloud. Table~\ref{tab: dataset} provides the statistics of these datasets. Specifically, the BGL (Blue Gene/L) dataset is a supercomputing system log dataset collected by Lawrence Livermore National Labs (LLNL)~\cite{oliner2007supercomputers, he2020loghub}. The Thunderbird dataset originates from a Thunderbird supercomputer at Sandia National Labs (SNL)~\cite{oliner2007supercomputers, he2020loghub}. Although existing studies often use 10 million continuous lines from the Thunderbird dataset for evaluation~\cite{le2022log, le2021log}, they do not specify which 10 million logs they employed; therefore, we use the first 10 million logs in this study.
The industrial dataset is collected from the production environment of \cloud between 01/02/2023 to 01/03/2023, generated by 20 internal microservices providing the abilities of API Gateway, user management, auto-scaling, load balance, etc. We manually identify and label the anomalies in the log messages based on the diagnostic reports provided by our site reliability engineering (SRE) team. These reports document the starting times of actual system issues, which allows us to manually identify the anomalous log messages.
After a thorough labeling process, we obtained a log data dataset with approximately 1.5 million lines of log messages, covering 60 types of system problems such as service unavailability, incorrect request parameters, and null pointers
Although the obtained number of log lines is smaller than that of the BGL and Thunderbird datasets due to the extensive manual labeling effort, our dataset contains a significantly higher number of templates, showcasing its diversity.

\subsubsection{Log Message Grouping}\label{sec: log_message_grouping}
It is a common practice to group log messages into windows and decide whether each window is anomalous or not, \eg grouping log messages within fixed or sliding time windows\cite{le2022log}.
For the industry dataset, we use short sliding time windows, specifically a ten-minute time window with a step size of two minutes. We choose this approach to replicate real-world scenarios that require real-time anomaly detection. 
The use of a short sliding time window has two main advantages. First, it ensures that anomalies are promptly detected and conserves memory.
Second, there should be some overlap between two adjacent windows to ensure that anomalies spanning multiple windows are not missed.
Furthermore, to avoid data leakage, we keep the time windows in chronological order without shuffling.
The recent empirical study~\cite{le2022log} has demonstrated that the length of the window size and whether or not to shuffle the windows can significantly impact the performance of log-based anomaly detection. 
Therefore, to ensure consistency with existing studies, we use the same window setting for the public dataset as in \cite{le2022log}, namely, using a fixed window size of one hour without overlapping.

\subsubsection{Implementation Details}
We implemented \nm as a microservice with approximately 1000 lines of Python code for easy use in \cloud. We conducted the following experiments on a Linux server running Ubuntu 18.04.6 LTS with 32GB of RAM and an Intel(R) Xeon(R) Platinum 8268 CPU @ 2.90GHz.
We set the hyperparameters empirically and fixed them as they were found to generalize well across different datasets without tuning. For internal node traversal, we set the length of the token key, i.e., $K=3$, and the maximum number of child nodes $c_{max}=3$. For leaf node update, we set the matching similarity threshold $\theta_{match}=0.5$. For anomaly detection, we fixed the temperature $\tau$ in Equation~\ref{equ: anomaly_score} to 10.

\begin{table}[]
\centering
\small
\caption{Statistics of evaluation datasets}
\label{tab: dataset}
\begin{tabular}{ccccc}
\toprule
\textbf{Dataset} & \textbf{BGL} & \textbf{Thunderbird} & \textbf{Industry} \\ \midrule
\# Log messages & 4,747,963 & 10,000,000 & 1,488,073 \\ \midrule
\# Templates & 456 & 1,504 & 3,241 \\ \midrule
\begin{tabular}[c]{@{}l@{}}\# Train windows\\ (anomaly ratio)\end{tabular} & \begin{tabular}[c]{@{}c@{}}2,884\\ (21\%)\end{tabular} & \begin{tabular}[c]{@{}c@{}}416\\ (55\%)\end{tabular} & \begin{tabular}[c]{@{}c@{}}3,048\\ (13\%)\end{tabular} \\ \midrule
\begin{tabular}[c]{@{}l@{}}\# Test windows\\ (anomaly ratio)\end{tabular} & \begin{tabular}[c]{@{}c@{}}722\\ (24\%)\end{tabular} & \begin{tabular}[c]{@{}c@{}}105\\ (30\%)\end{tabular} & \begin{tabular}[c]{@{}c@{}}933\\ (18\%)\end{tabular} \\ \bottomrule
\end{tabular}
\vspace{-12pt}
\end{table}

\subsubsection{Evaluation Metrics}
For each window, we calculate four measures:
TP (true positive), which is the number of correctly predicted anomaly windows; FP (false positive), which is the number of predicted anomaly windows that are actually normal; TN (true negative), which is the number of correctly predicted normal windows; and FN (false negative), which is the number of predicted normal windows that are actually anomalous.
Based on these numbers, we calculate precision (PC)=$\frac{TP}{TP+FP}$ and recall (RC) =$\frac{TP}{TP+FN}$. We also use F1 scores=$\frac{2\times PC \times RC}{PC+RC}$ to evaluate the overall performance.

\begin{table*}[]
\small
\captionsetup{justification=centering}
\centering
\caption{Experimental results of offline anomaly detection\\ (We use $*$ to denote unsupervised methods, others are supervised ones)}
\vspace{-12pt}
\label{tab: offline_setting}
\begin{tabular}{cccccccccc}
\toprule
\multirow{2}{*}{Method} & \multicolumn{3}{c}{BGL} & \multicolumn{3}{c}{Thunderbird} & \multicolumn{3}{c}{Industry} \\
 & Precision & Recall & \textbf{F1 score} & Precision & Recall & \textbf{F1 score} & Precision & Recall & \textbf{F1 score} \\ \midrule
IF$^*$ & 0.125 & 0.615 & 0.208 & 0.291 & 0.968 & 0.448 & 0.176 & 0.994 & 0.299 \\
DT & 1.000 & 0.570 & 0.726 & 1.000 & 0.893 & \underline{0.912} & 0.942 & 0.788 & 0.858 \\ 
LR & 0.738 & 0.437 & 0.549 & 0.842 & 0.516 & 0.640 & 0.818 & 0.655 & 0.727 \\
\midrule
DeepLog$^*$ & 0.241 & 0.895 & 0.380 & 0.295 & 1.000 & 0.456 & 0.358 & 0.909 & 0.513 \\
LogAnomaly$^*$ & 0.268 & 0.862 & 0.409 & 0.307 & 1.000 & 0.470 & 0.360 & 0.927 & 0.519 \\
RobustLog & 0.942 & 0.961 & \ \underline{0.951} & 1.000 & 0.710 & 0.830 & 0.984 & 0.764 & 0.860 \\
NeuralLog & 0.881 & 0.886 & 0.883 & 0.713 & 0.719 & 0.715 & 0.889 & 0.895 & \underline{0.887} \\ \midrule
\nm & 0.993 & 0.993 & \textbf{0.990} & 1.000 & 0.903 & \textbf{0.949} &  1.000 & 0.931 & \textbf{0.908} \\
\bottomrule
\end{tabular}
\vspace{-12pt}
\end{table*}

\subsubsection{Comparative Methods}
We have selected the following state-of-the-art studies as our comparative methods:
\begin{itemize}[leftmargin=*, topsep=0pt]
\item \textbf{Loglizer (IF/LR/DT)}~\cite{he2016experience} is a log-based anomaly detection library encompassing a variety of machine learning (ML) methods. The library processes parsed log messages as input and computes a count feature vector representing the distribution of template counts. This count vector is subsequently used as input for ML-based algorithms to perform anomaly detection. Prominent ML algorithms within Loglizer include isolation forest (IF), linear regression (LR), and decision tree (DT).

\item \textbf{DeepLog}
\cite{du2017deeplog} employs a Long Short-Term Memory (LSTM) model as its core component. DeepLog takes windows of log messages as input and predicts the next log event. Anomalies are detected if the prediction differs from the actual event.
\item \textbf{LogAnomaly}~\cite{meng2019loganomaly} aims to detect anomalies by combining log count vectors and log semantic vectors, which also utilizes a forecasting-based mechanism to reflect anomalies.
\item \textbf{LogRobust}~\cite{zhang2019robust} aims to capture the semantics of log messages through word embeddings to address the issue of ever-changing log messages. Each log message is encoded as a representation based on word vectors, which is modeled by an attention-based Bidirectional LSTM and trained in a supervised manner.
\item \textbf{NeuralLog}~\cite{le2021log} aims to bypass the log parsing step; instead, NeuralLog extracts semantic meaning from raw log messages and represents them as vectors to detect anomalies using a Transformer-based classification model. 
\end{itemize}


\subsection{Effectiveness under Offline Setting (RQ1)}\label{sec: exp_offline}
\noindent In this RQ, we evaluate the effectiveness of \nm in an offline setting, where we train and test the models with the previously collected data. 
To simulate this process, we split the whole dataset in chronological order, with the first 80\% data for training and the remaining 20\% data for testing, and keep them fixed following previous studies~\cite{le2022log,le2021log}. 
The statistics of training and testing data are shown in Table~\ref{tab: dataset}. 
We incorporate the labels as feedback when training \nm, and we do not use any feedback during testing, which is consistent with supervised models.

The evaluation results are presented in Table~\ref{tab: offline_setting}, which categorizes the baseline methods into ML-based (upper half) and DL-based (lower half) approaches. The highest F1 score is emphasized in \textbf{bold}, and the second-best score is \underline{underlined}. 
We can make the following observations: 
(1) Unsupervised methods (IF, DeepLog, and LogAnomaly) show significantly lower effectiveness than supervised approaches in terms of  F1 score. The reason is that they are originally designed to consider all training windows as normal. Therefore, during inference, they classify every window containing unseen log templates as anomalies, resulting in high recall but low precision.
(2) Supervised methods achieve a better balance between recall and precision by learning from labeled data. Furthermore, with sufficient labels, ML-based methods (i.e., DT) can still achieve comparable or better performance (on Thunderbird) compared to DL-based methods. This observation suggests that the count distributions of different templates could facilitate anomaly detection.
The two observations (1) and (2) are consistent with the recent empirical studies~\cite{le2022log,chen2021experience} that benchmark existing DL-based methods.
(3) \nm attains the highest F1 score across all three datasets, ranging from 0.908 to 0.990. This result demonstrates the effectiveness of detecting anomalies from count distribution (i.e., results of TDA), as well as utilizing the labeled data in the training data (i.e., experts' feedback).
{
\vspace{-5pt}
\begin{tcolorbox}[breakable,width=\linewidth-2pt,boxrule=0pt,top=1pt, bottom=0pt, left=1pt,right=1pt, colback=gray!20,colframe=gray!20]
\textbf{Answer to RQ1.} 
\nm achieves the best F1 score (0.990, 0.949 and 0.908) among all state-of-the-art baselines across two public datasets and one industrial dataset. The results demonstrate that \nm can effectively fuse the results of TDA and experts' feedback in the offline setting, thus meeting the requirement of being accurate for practical log-based anomaly detection.
\end{tcolorbox}
\vspace{-12pt}
}

\begin{figure}[t]
  \centering
  \mbox{
     \subfigure[BGL\label{fig: online_BGl}]{\includegraphics[width=0.47\columnwidth]{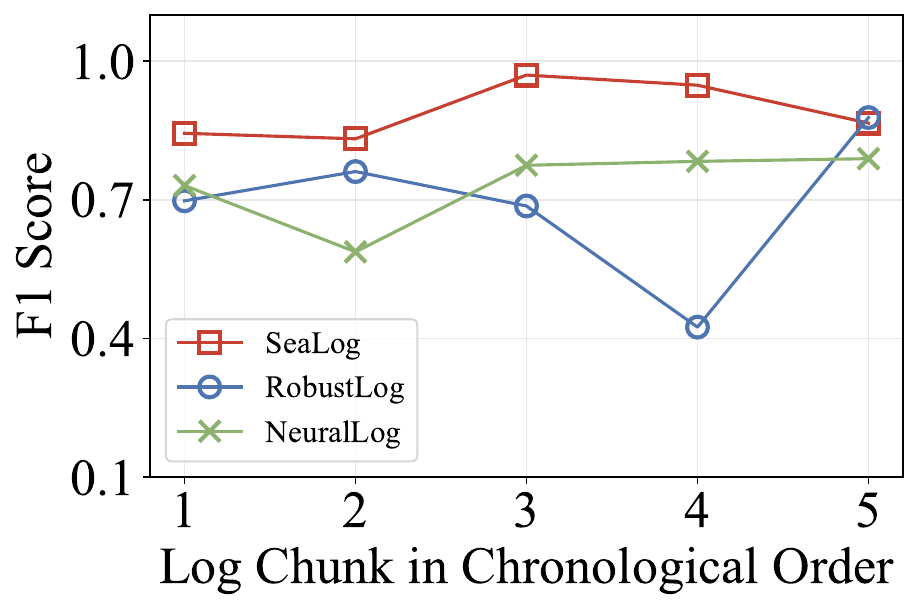}}\quad
    
    \subfigure[Industry 
    \label{fig: online_industry}]
    {\includegraphics[width=0.47\columnwidth]{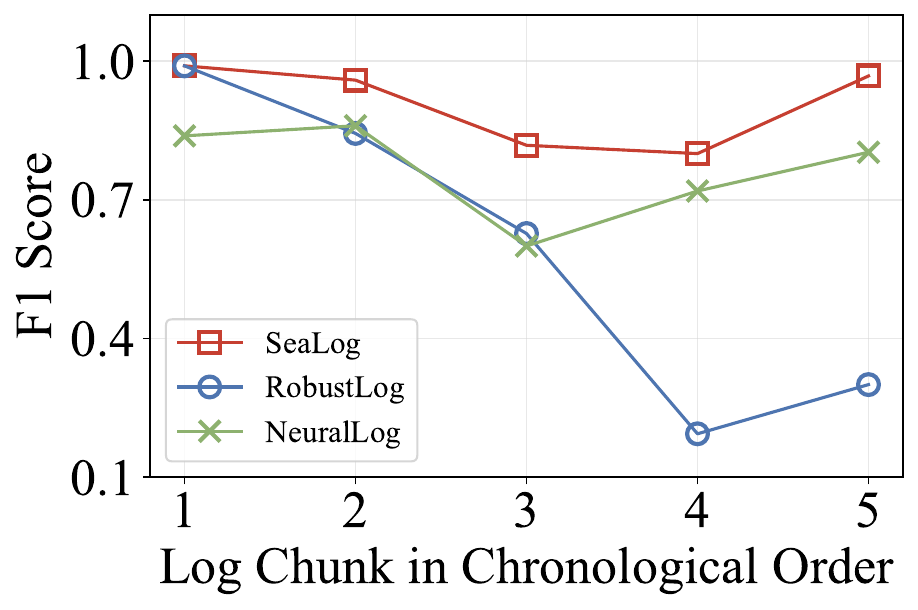}}
    }
    \vspace{-12pt}
  \caption{Experimental results of online anomaly detection}
  \label{fig: online_exp_all}
\vspace{-12pt}
\end{figure}

\subsection{Effectiveness under Online Setting (RQ2)}\label{sec: exp_online}
\noindent Our preliminary study has shown that real-world log data is continuously evolving, which highlights the need for a practical anomaly detection system that can adapt to these changes and maintain high accuracy. 
To address this requirement, we investigate the effectiveness of \nm within an online setting in RQ2.
Based on the results of RQ1, we select the most effective model (i.e., RobustLog and NeuralLog) as strong baselines for this question.
To simulate a scenario where new log messages are received, we divide the BGL and Industry datasets into six even chunks (numbered from 0 to 5) in chronological order, each containing log messages with different templates.
In practice, on-call engineers must verify suspicious logs reported by an anomaly detector, which naturally provides labeled data allowing us to continue training a model on-the-fly.
We train the models using the preceding chunk and evaluate the trained models on the subsequent chunk (e.g., training with chunk 0 and testing with chunk 1), ultimately yielding five evaluation results.
Although RobustLog and NeuralLog were not initially designed for processing streaming feedback, we equip them with the capability to perform continuous training by fitting the incoming chunk, enabling a fair comparison with \nm.

Figure~\ref{fig: online_exp_all} presents the experimental results. Our observations are as follows: 
(1) \nm consistently achieves the highest F1 scores across all chunks for both datasets, demonstrating its successful adaptation to new chunks and maintaining stable performance.
(2) RobustLog's performance significantly declines from chunk 1 to chunk 4 in both datasets, indicating that it experiences catastrophic forgetting~\cite{french1999catastrophic} when accommodating new chunks. Nevertheless, its performance improves in the 5th chunk on the BGL dataset, as chunks 4 and 5 share considerable data overlap, enabling RobustLog to detect the most recent anomalies. Conversely, the Industry dataset is more complex and has less overlap between chunks, resulting in substantially lower performance for RobustLog compared to other models.
(3) NeuralLog exhibits greater stability in performance than RobustLog due to its transformer architecture, which possesses a larger number of parameters capable of memorizing observed data and mitigating the catastrophic forgetting issue.

{
\begin{tcolorbox}[breakable,width=\linewidth-2pt,boxrule=0pt,top=1pt, bottom=0pt, left=1pt,right=1pt, colback=gray!20,colframe=gray!20]
\textbf{Answer to RQ2.} 
\nm consistently surpasses the strong baseline methods (RobustLog and NeuralLog) on both BGL and industry datasets within the online setting. The results demonstrate that \nm is able to adapt to evolving logs, thereby meeting the requirements of practical log-based anomaly detection that demands adaptability. 
\end{tcolorbox}
\vspace{-12pt}
}

\subsection{Impact of Experts' Feedback (RQ3)}\label{sec: exp_expert_feedback}
\noindent 
In this RQ, we aim to study how the number of queries affects \nm's effectiveness.
To achieve this goal, we train \nm with only 50\% of the training data for BGL and industry datasets, which mimics the scenario that the model might not be well trained initially.
Then, we use the same test set as in RQ1 to evaluate \nm's performance.
While testing, we vary the amount of feedback given to \nm by tuning the threshold to issue a query, \ie $\theta_{query}$ in the range of [0, 1] with step size of 0.1.

\begin{figure}[t]
  \centering
  \mbox{
     \subfigure[BGL\label{fig: feedback_bgl}]{\includegraphics[width=0.51\columnwidth]{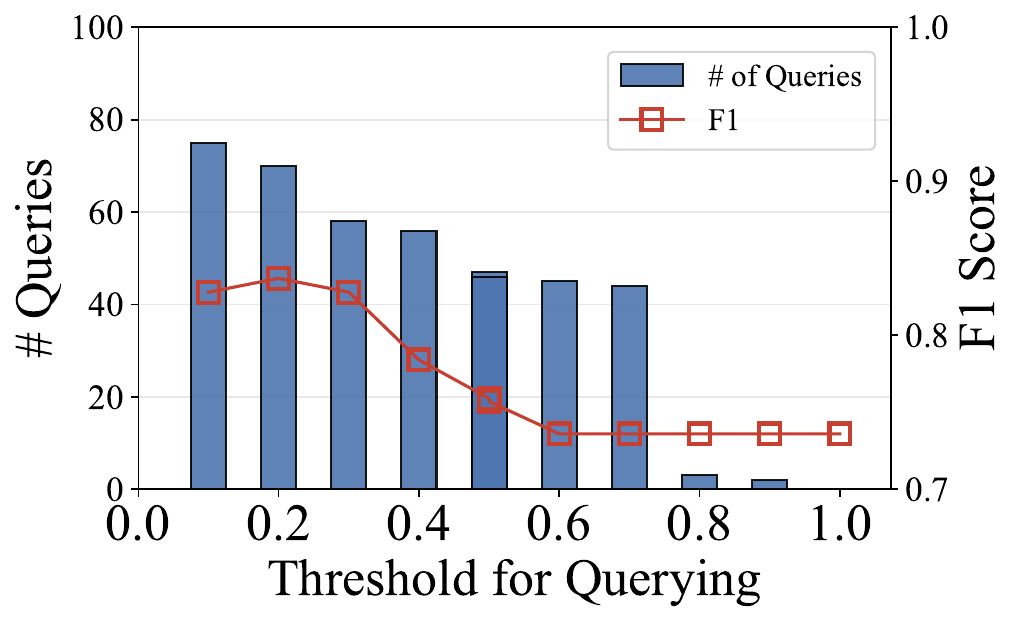}}\hspace{-5pt}\quad
    
    \subfigure[Industry\label{fig: feedback_industry}]
    {\includegraphics[width=0.51\columnwidth]{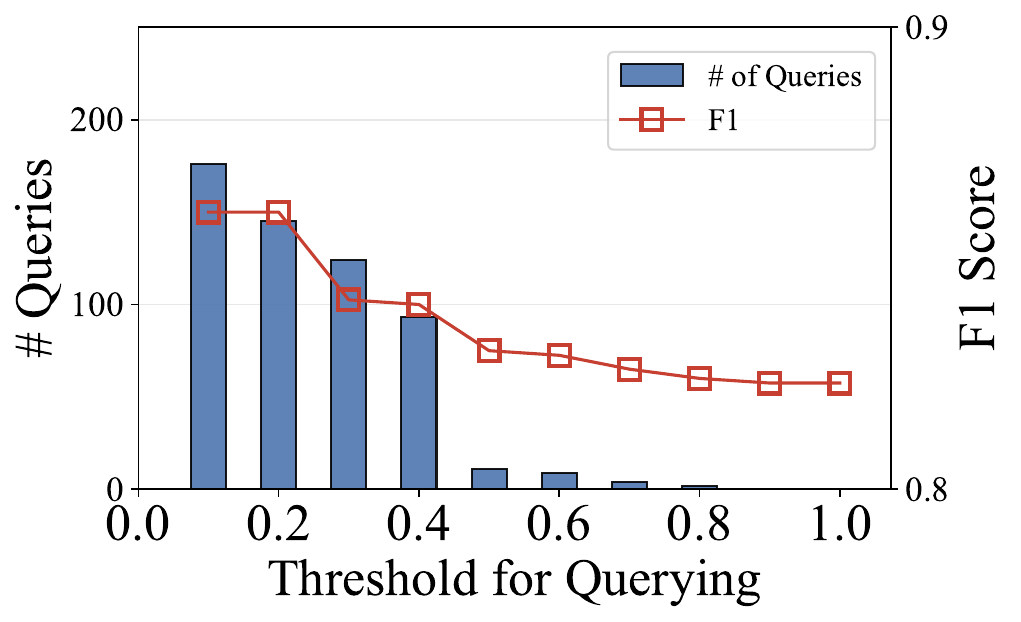}}
    }
  \vspace{-12pt}
  \caption{Anomaly Detection performance w.r.t. the number of queries to experts}
  \label{fig: feedback}
\vspace{-12pt}
\end{figure}

The results are presented in Figure~\ref{fig: feedback}. 
We can obtain the following observations: 
(1) The expert is asked less when increasing the threshold $\theta_{query}$. With less feedback, the performance of \nm is degraded. However, even if no feedback is taken (i.e., setting $\theta_{query}=0$), \nm can still achieve around 0.73 F1 score for BGL and 0.82 for industry data, respectively. In this setting, anomaly detection is only done with the trained TDA.
(2) When the query threshold $\theta_{query}$ is set to values greater than or equal to 0.8 and 0.5 for the BGL and industry datasets respectively, the number of queries reduces significantly. This indicates that there are only a few samples in the datasets with anomaly scores that exceed these threshold values. 
The reason for this is that the EVT-based anomaly detection module in \nm tends to conservatively assign high anomaly scores to a small number of anomalous samples.
(3) After being trained, TDA only queries fewer than 80 and 190 times on BGL and Industry datasets, respectively.  For the industry dataset, it indicates most 6 queries per day can me received from TDA.
This indicates that TDA can effectively filter out most of the templates and only interact conservatively with experts.

{
\vspace{-5pt}
\begin{tcolorbox}[breakable,width=\linewidth-2pt,boxrule=0pt,top=1pt, bottom=0pt, left=1pt,right=1pt, colback=gray!20,colframe=gray!20]
\textbf{Answer to RQ3.} 
The performance of \nm improves as the amount of feedback increases. Owing to the EVT-based anomaly detection model in TDA, most normal log messages are filtered out, resulting in only 190 queries in total (6 queries per day) for the industry dataset. Furthermore, even without any feedback, the trained \nm maintains F1 scores of 0.71 and 0.82 on the BGL and Industry datasets, respectively.
\end{tcolorbox}
\vspace{-12pt}
}

\subsection{Time and Space Efficiency (RQ4)}\label{sec: exp_time_space}
\noindent T
A practical log-based anomaly detector must be capable of processing a large number of log messages efficiently, while also minimizing memory usage. In this research question, we compare \nm with baseline methods in terms of time and space efficiency. In this RQ, we only evaluate the TDA part because TDA is responsible for online log analysis, while the expert can be deployed remotely.

\textit{(1) Time efficiency} We measure the time required to perform anomaly detection on the complete BGL test set with the aim of simulating the scenario of deploying these methods in a production environment. 
For DL-based methods, we set them to evaluation mode without backpropagation. Additionally, since online-generated logs are not pre-parsed, we include the time required to parse raw log messages using Drain~\cite{he2017drain} for baseline methods. The time required for \nm encompasses all steps from preprocessing to anomaly detection. Note that I/O time is excluded for all methods, and these methods are executed without GPU acceleration, which may not be available when resources are constrained.

\begin{figure}[t]
  \centering
  \mbox{
     \subfigure[Time Efficiency Comparison\label{fig: time_efficiency}]{\includegraphics[width=0.49\columnwidth]{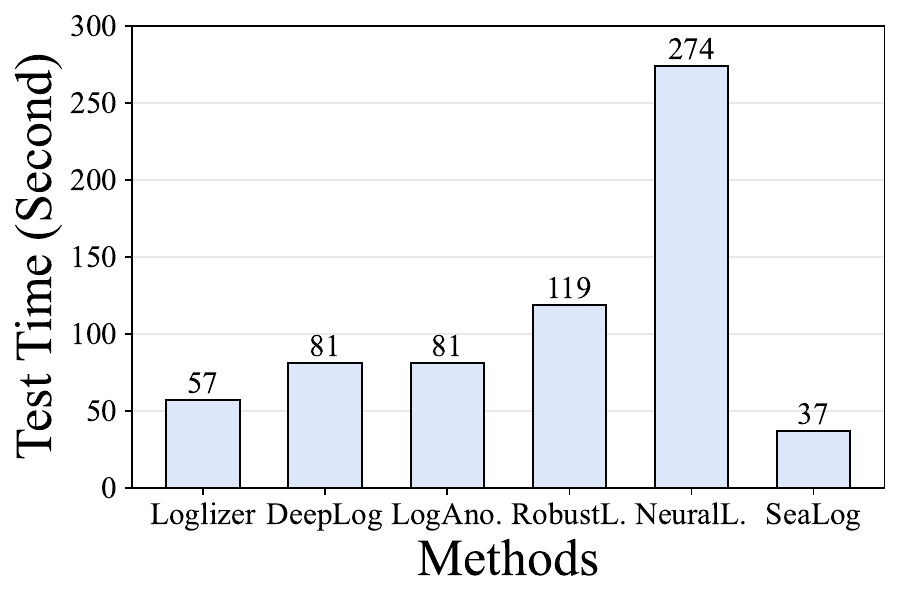}}\quad
    
    \subfigure[Space Efficiency Comparison 
    \label{fig: space_efficiency}]
    {\includegraphics[width=0.49\columnwidth]{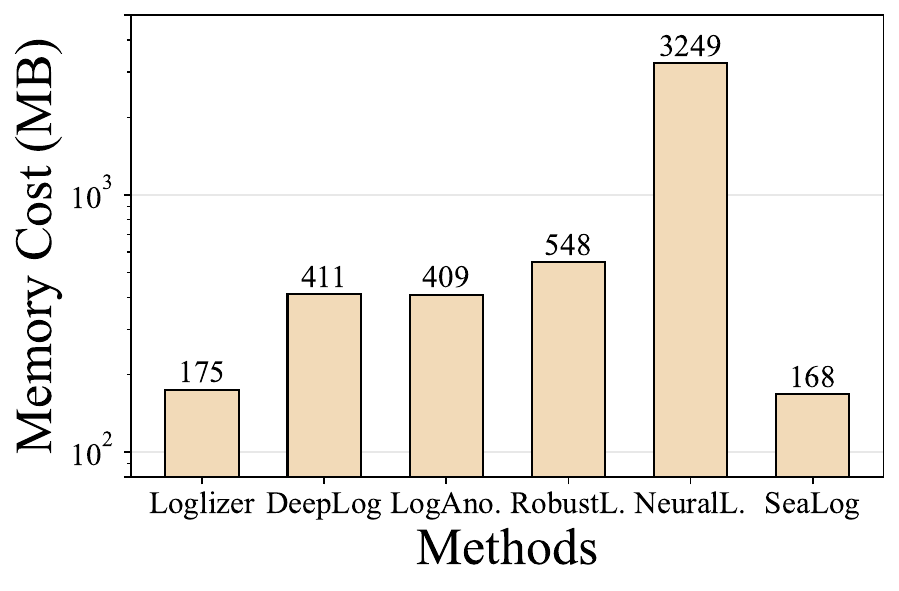}}
    }
   \vspace{-12pt}
  \caption{Experimental results of time and space efficiency comparison}
\vspace{-12pt}
\end{figure}

Figure~\ref{fig: time_efficiency} presents the comparison results. Since LR, IF, and DT present similar efficiency, their values are averaged and denoted as Loglizer in the figure. We can observe that: (1) \nm is the most efficient method, being 2 to 10 times faster than the baseline methods; (2) ML-based methods (Loglizer) demonstrate higher efficiency than DL-based methods but cost more time than \nm. The reason is that Loglizer requires a separate parsing step while \nm conducts log parsing and anomaly detection simultaneously; (3) DeepLog and LogAnomaly exhibit comparable performance, as they both utilize LSTM-based architectures. In contrast, RobustLog and NeuralLog require significantly more time than others, as they adopt networks with a greater number of parameters.

\textit{(2) Space efficiency}
We employ Memory Profiler~\cite{mem_profiler} to track the memory consumption of each method while processing a single log session. Since these methods operate on a single session during online log analysis, we chose one session for the comparison. 
The comparison emphasizes the maximum memory occupied by each method, as an out-of-memory error may occur if the available memory in an instance is less than what a method requires. Figure~\ref{fig: space_efficiency} displays the comparison results. 
The memory usage trend is similar to the time efficiency trend.
\nm demands the least memory during log processing, necessitating only 5\% to 41\% of the memory consumed by DL-based methods. This is primarily because \nm only needs to store the trie structure in memory, which is typically sparse. Loglizer demonstrates comparable space efficiency to \nm, since it only stores the count vector of a log session, and the subsequent ML-based method (e.g., LR) is memory-efficient.
DL-based methods necessitate storing a large number of parameters (e.g., network parameters), resulting in higher memory usage. NeuralLog, which adopts a heavy transformer architecture, consumes the most memory among the methods.

{
\begin{tcolorbox}[breakable,width=\linewidth-2pt,boxrule=0pt,top=1pt, bottom=0pt, left=1pt,right=1pt, colback=gray!20,colframe=gray!20]
\textbf{Answer to RQ4.} 
\nm outperforms the selected state-of-the-art baseline methods, being 2 to 10 times faster, and requires only 5\% to 41\% of the memory consumed by recent deep learning-based methods.
The results demonstrate that \nm is efficient concerning both time and space efficiency, thus meeting the requirement of being lightweight for practical anomaly detection.
\end{tcolorbox}
\vspace{-12pt}
}

\subsection{Case Study}\label{sec: case_study}
\begin{figure}[t]
  \centering
  \mbox{
     \subfigure[Prompt design for anomaly detection\label{fig: gpt_prompt}]{\includegraphics[width=0.51\columnwidth]{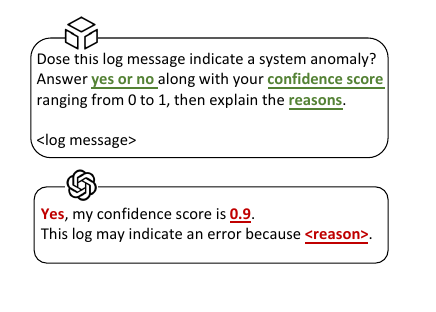}}\hspace{-5pt}\quad
    
    \subfigure[Probability distribution of ChatGPT\label{fig: gpt_prob_distribution}]
    {\includegraphics[width=0.5\columnwidth]{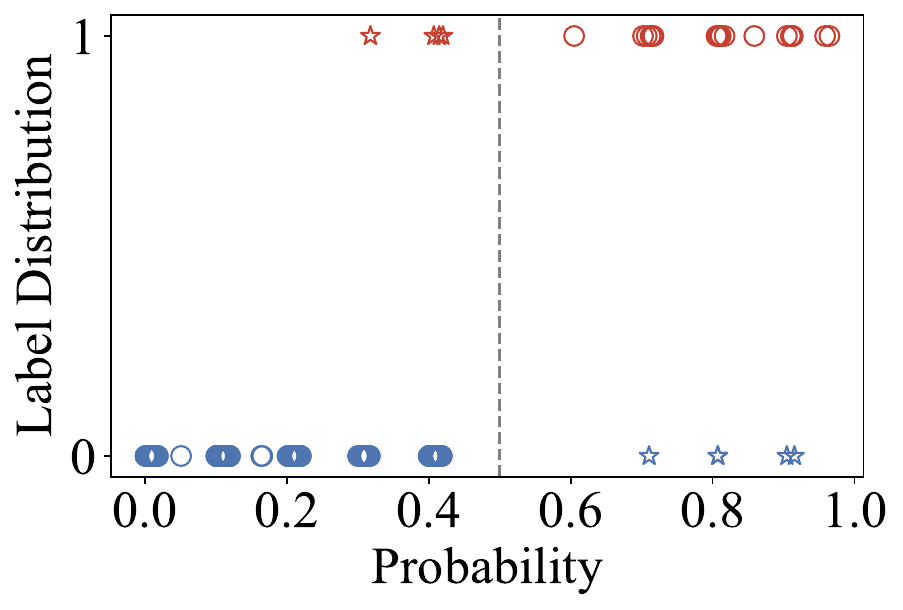}}
    }
   \vspace{-12pt}
  \caption{A case study of using ChatGPT as an expert}
  \label{fig: case-study}
\vspace{-12pt}
\end{figure}

In this section, we study the feasibility of employing ChatGPT as an expert to support on-call engineers in the loop of \nm. 
Specifically, we aim to measure the extent to which ChatGPT can provide feedback that is as consistent as that of on-call engineers.
To achieve this, compare the decisions of ChatGPT and that of experts (\ie groundtruth) for querying the same set of logs.
We randomly select 200 lines of log messages from the production environment of \cloud, 24 of which indicate a system problem, with the rest being normal logs. We manually remove sensitive information in the logs for privacy issues. 
The prompt to query ChatGPT is shown in Figure~\ref{fig: gpt_prompt}, which requires ChatGPT to output a confidence score, along with reasons for its decision, which could assist on-call engineers in understanding its output. Note that we have made the queries and responses publicly available in our Github repository.

By comparing the ``yes or no'' answers of ChatGPT with human labels, we can obtain an f1 score of 0.816, recall of 0.833 and precision of 0.8. Figure~\ref{fig: gpt_prob_distribution} shows the detailed probability distribution of ChatGPT's output.
We observed that most cases were appropriately classified with a 0.5 threshold. However, some false predictions were still present (represented as stars in the figure).
After manually inspecting these false predictions, we noticed that (1) ChatGPT may miss some anomalies if log messages lack evident semantics that indicate an error. 
(2) the normal logs that are wrongly classified as abnormal were subtle anomalies that engineers typically disregard, such as "failed to login, wrong password." Such logs often contain keywords such as "failed," indicating abnormal semantics.

These observations suggest that ChatGPT is powerful in identifying anomalies purely based on the semantics of log messages. It is promising to integrate ChatGPT within the framework of \nm, as evidenced by its high consistency with human decisions (\ie high F1 score of 0.816). However, integration with more domain knowledge of on-call engineers could further enhance its performance, such as prompt designs that enable ChatGPT to ignore login errors in this case.
Since our paper's primary focus is not designing a high-performing expert, we leave this for future work. 

\section{Threats to Validity}
We identify the following threats to external and internal validity.

\textbf{External validity.} 
The type of log-based anomaly is the first external threat. \nm assumes that log templates with low occurrence counts are more likely to be anomalous. While this assumption is generally valid based on our experience and experimental results (e.g., RQ1 and RQ2), there are cases where anomalous log templates can burst within a short period, leading to a high occurrence count. Nevertheless, such scenarios typically indicate a severe system problem and can be easily identified by on-call engineers via other measurements, such as the key performance indicators (e.g., heartbeat).
The second external threat is the study's object. Our motivation for a lightweight and adaptive method is driven by real-world scenarios in a single cloud system, namely \cloud. However, \cloud is a representative world-leading cloud provider with a vast scale. Besides, we adopt public datasets for evaluation. Hence, the evaluation results should be representative and compelling.


\textbf{Internal validity.} Implementation and parameter settings are the primary internal threats to the validity of our study. To mitigate these threats, we have implemented several measures.
First, we have employed peer code review while implementing \nm, and for the baseline methods, we have utilized open-sourced code released by the original paper or highly-rated replications on Github.
Second, we have used the hyperparameters recommended by the original authors wherever available. If the hyperparameters were not available, we performed an intensive hyperparameter tuning process, such as grid search, to optimize the models for baseline methods.
Third, to ensure the reproducibility of our results, we have made our code and partial data publicly available.

\section{Related Work}
\noindent Prior research related to log-based anomaly detection can be broadly categorized into two classes, \textit{machine learning (ML)-based} methods and \textit{deep learning (DL)-based} methods.

\textbf{Machine learning-based methods} include the use of principal component analysis (PCA) by Xu et al.\cite{xu2009largescale} for mining system problems from console logs and the work of Lou et al.\cite{lou2010mining}, who detects system anomalies by mining invariants among log messages. Lin et al.\cite{lin2016log} proposed LogCluster, which recommends representative log sequences for problem identification by clustering similar log sequences. He et al.\cite{he2018identifying} proposed Log3C, which incorporates system monitoring metrics into the identification of high-impact issues in service systems. Loglizer~\cite{he2016experience} provides a comprehensive evaluation of using ML-based methods for log-based anomaly detection.

These ML-based methods generally require fewer computation resources for execution, however, they are not tailored for addressing evolving logs as in a complex cloud system. Differently, \nm achieves better efficiency due to our lightweight design and is capable of adapting ever-changing logs.

\textbf{Deep learning-based methods} for log-based anomaly detection have been approached through various methods. One such method, proposed by Du et al.\cite{du2017deeplog}, is Deeplog, which utilizes a Long Short-Term Memory (LSTM) network to model log sequences. LogAnomaly\cite{meng2019loganomaly} further utilizes log count vectors and log semantic vectors to model log sequences more comprehensively. However, both Deeplog and LogAnomaly are trained in an unsupervised manner, which has been shown to be less effective than supervised models~\cite{le2022log}. Typical supervised solutions include a CNN-based approach~\cite{lu2018detecting} and GRU-based LogRobust~\cite{zhang2019robust}. Since labeled data is usually insufficient due to the labor-intensive nature of manual labeling, the semi-supervised method, PLElog~\cite{yang2021semi}, addresses this problem via label probabilistic estimation.

DL-based methods utilize different neural network structures (\ie LSTM and Transformers) to capture patterns from historical log messages. However, these structures are too complex in terms of time and space complexity to be deployed locally in an instance. Additionally, these methods still encounter accuracy degradation when logs change. In contrast, \nm utilizes a TDA for local and efficient anomaly detection, which is continuously improved by a remote expert.

\section{Conclusion}
In this paper, we presented \nm, a scalable and adaptive log-based anomaly detection framework designed to meet the practical requirements of accuracy, lightweight design, and adaptiveness in cloud systems.
\nm utilizes a trie-based detection agent (TDA) for lightweight and adaptive anomaly detection in a streaming manner. We also incorporate expert feedback, including utilizing large language models (LLMs) as an expert, to continuously enhance the system's accuracy.
Experimental results on two public datasets and an industrial dataset from CloudX showed that \nm is effective, achieving F1 scores between 0.908 and 0.990. Moreover, \nm maintained high and consistent accuracy, even in the presence of evolving log data, while being 2 to 10$\times$ faster and requiring 5\% to 41\% memory resources compared to existing methods.
Overall, our proposed \nm framework provides a practical and efficient solution for log-based anomaly detection in cloud systems, while highlighting the significance of combining human expertise and machine learning techniques for improved performance.


\balance
\bibliographystyle{ACM-Reference-Format}
\bibliography{ICSE24}
\end{document}